\documentclass[runningheads]{llncs}
\usepackage[T1]{fontenc}
\usepackage{graphicx}
\usepackage{hyperref}
\usepackage{multirow}
\usepackage[most]{tcolorbox}
\usepackage{lipsum} 
\usepackage{pdfpages}
\DeclareUnicodeCharacter{2212}{-}

\begin{document}
\title{A Zero-Shot LLM Framework for Automatic Assignment Grading in Higher Education}
\titlerunning{Automatic Assignment Grading in Higher Education}
%
\author{
Calvin Yeung\inst{1,3} \and
Jeff Yu\inst{1} \and 
King Chau Cheung \inst{1} \and
Tat Wing Wong \inst{1} \and \\
Chun Man Chan \inst{1} \and
Kin Chi Wong \inst{2} \and 
Keisuke Fujii \inst{3}
 }

\authorrunning{C. Yeung et al.}
%
\institute{Department of Statistics, The Chinese University of Hong Kong, Hong Kong, China 
\email{\{kingchaucheung,tw.wong,chunmanchan\}@cuhk.edu.hk}\\
\and
Centre for Learning Enhancement And Research, The Chinese University of Hong Kong, Hong Kong, China \\
\email{\{kc.wong\}@cuhk.edu.hk}\\
\and
Graduate School of Informatics, Nagoya University, Nagoya, Japan\\
\email{\{yeung.chikwong,fujii\}@g.sp.m.is.nagoya-u.ac.jp}
}


%
\maketitle              
\begin{abstract}
Automated grading has become an essential tool in education technology due to its ability to efficiently assess large volumes of student work, provide consistent and unbiased evaluations, and deliver immediate feedback to enhance learning. However, current systems face significant limitations, including the need for large datasets in few-shot learning methods, a lack of personalized and actionable feedback, and an overemphasis on benchmark performance rather than student experience.
To address these challenges, we propose a Zero-Shot Large Language Model (LLM)-Based Automated Assignment Grading (AAG) system. This framework leverages prompt engineering to evaluate both computational and explanatory student responses without requiring additional training or fine-tuning. The AAG system delivers tailored feedback that highlights individual strengths and areas for improvement, thereby enhancing student learning outcomes.
Our study demonstrates the system's effectiveness through comprehensive evaluations, including survey responses from higher education students that indicate significant improvements in motivation, understanding, and preparedness compared to traditional grading methods. The results validate the AAG system's potential to transform educational assessment by prioritizing learning experiences and providing scalable, high-quality feedback.

\end{abstract}

\section{Introduction}
In the field of education technology, automated grading has long been a key objective due to its ability to efficiently assess large volumes of work \cite{burrows2005management}, deliver consistent and unbiased evaluations \cite{haley2007measuring}, and provide immediate feedback to enhance student learning \cite{hirschman2000automated}. It also allows educators to focus on developing engaging learning experiences and leverage data-driven insights for improved teaching strategies.

Automated grading systems have evolved significantly since their inception in the 1960s, initially focusing on programming and essay evaluation \cite{hollingsworth1960automatic,page1966imminence}. Traditional methods, typically compare student submissions with reference answers using text similarity and predefined rules \cite{heilman2013ets,liu2019automatic,messer2024automated,ureel2019automated}.  Recent developments in Large Language Models (LLMs) have introduced new opportunities for automated grading. LLMs such as BERT \cite{devlin2018bert}, GPT-3 \cite{brown2020language}, and GPT-4 \cite{achiam2023gpt} have shown remarkable capabilities in tasks like question answering and summarization. Where few-shot learning approaches (see Section \ref{sec:related_work} for more details) have become common in automated grading systems. 

Despite significant advancements, current automated grading systems still face notable challenges. Firstly, few-shot learning methods demand substantial datasets, which are difficult to acquire for dynamic or specialized course content. Building extensive labeled datasets for each possible assignment or course content is time-consuming and often impractical, especially in fast-evolving subjects. Zero-shot learning, in contrast, doesn't require such extensive datasets and can be generalized across many tasks without additional task-specific training. 

Therefore zero-shot approaches were more scalable and adaptable, particularly for grading diverse assignments or subjects without the need for retraining. Additionally, previous efforts \cite{kortemeyer2023can} to implement zero-shot learning in grading have shown limited effectiveness, though there is potential for improvement through prompt engineering, an area that remains underexplored. Secondly, these systems often lack a strong emphasis on providing personalized, actionable feedback to students. Lastly, evaluations tend to prioritize benchmark performance over student experience and the practical value delivered to learners.

\begin{figure}
\includegraphics[width=\textwidth]{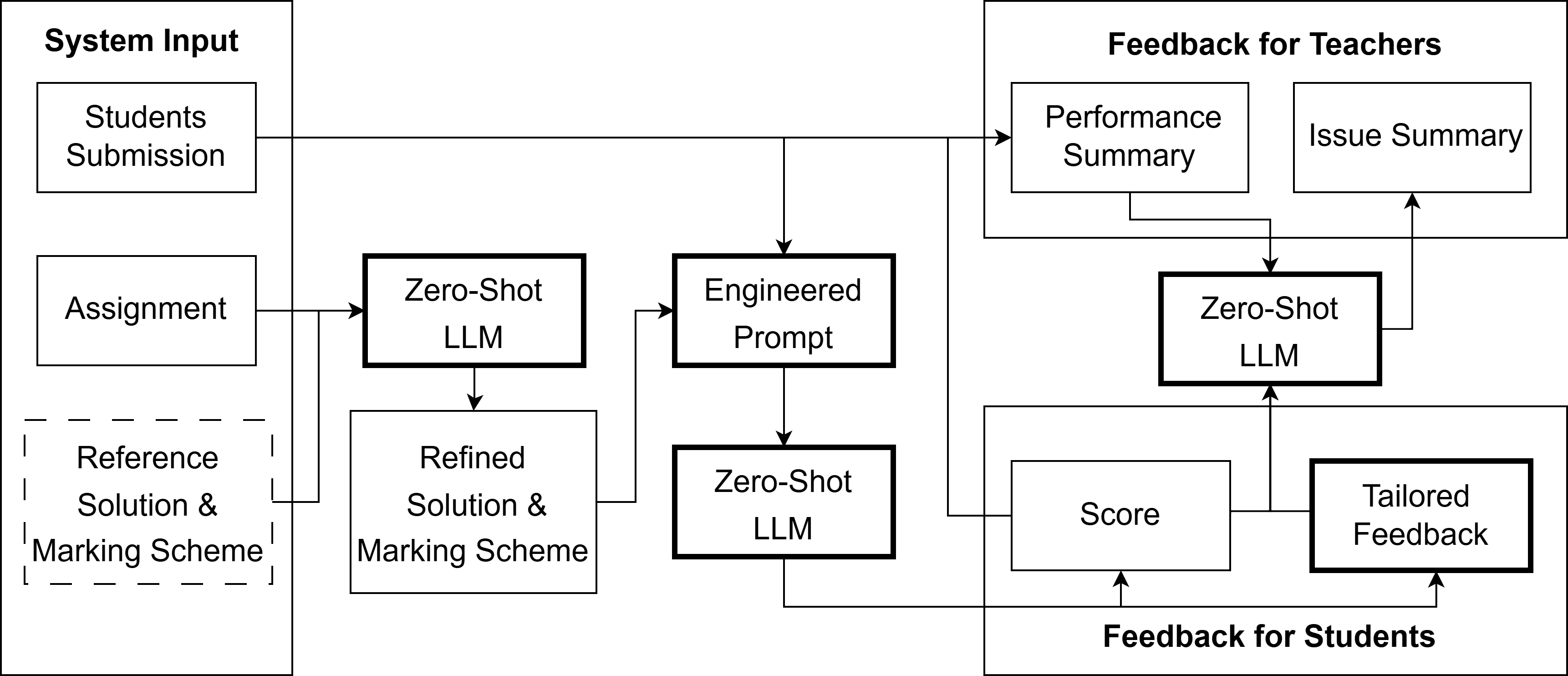}
\caption{Overview of the Zero-Shot LLM-Based AAG System. The bolded box highlights the core innovation of this study: an AAG system integrating prompt engineering, zero-shot LLM capabilities, and tailored feedback for students. The dashed line box indicates non-complementary input.} \label{fig1}
\end{figure}

To address these challenges, we propose a Zero-Shot LLM-Based Automated Assignment Grading (AAG) system, illustrated in Fig. \ref{fig1},  which leverages prompt engineering to evaluate student submissions without requiring additional training or fine-tuning. This system is capable of evaluating both computational and explanatory responses while providing personalized feedback. The feedback not only highlights mistakes but also provides tailored and actionable suggestions for improvement. 
Our evaluation approach goes beyond simply assessing grading accuracy, focusing on how the approach enhances learning experiences for both undergraduate and graduate students, based on their survey responses. The contributions of this paper are as follows:
\begin{enumerate}
    \item \textbf{Zero-shot LLM framework for AAG}: The proposed framework demonstrates effective assignment evaluation via prompt engineering without requiring additional training or fine-tuning on specific datasets.

    \item \textbf{Enhanced Learning through Tailored Feedback}: By delivering actionable, personalized feedback, the system improves individual learning outcomes, helping students identify strengths and address weaknesses.

    \item \textbf{Impact Validated via Student-Centered Metrics}: Through surveys and evaluations, the framework's significant role in enhancing student engagement, understanding, and preparedness is demonstrated, providing clear evidence of its practical educational value.
\end{enumerate}

\section{Related Work}
\label{sec:related_work}

\textbf{Automated Grading Systems}. 
Automated grading tools were first introduced in the 1960s for programming \cite{hollingsworth1960automatic} and essay evaluation \cite{page1966imminence}. Since then, numerous methods have been developed, including unit testing \cite{messer2024automated}, rule-based approaches \cite{liu2019automatic,ureel2019automated}, and techniques based on stacking and domain adaptation \cite{heilman2013ets}. These methods typically compare student submissions with reference answers using text similarity, measurable text characteristics (e.g., sentence length, essay length, number of prepositions, and punctuation), and predefined rules. However, with the rapid advancement of deep learning, machine learning, and neural network-based methods have proven more effective than traditional non-neural approaches \cite{riordan2017investigating}. Consequently, various neural models have been applied to automated grading, including Long Short-Term Memory (LSTM) networks \cite{mizumoto2019analytic,riordan2017investigating}, Convolutional Neural Networks (CNNs) \cite{shin2021more}, and Prototypical Neural Networks \cite{zeng2023generalizable}. Additionally, additive model-based methods \cite{condor2024explainable} have been introduced to enhance model explainability.

\textbf{Large Language Models in Automated Grading Systems}. 
In recent years, the significant growth of LLMs in natural language processing has led to their adoption in automated grading systems for evaluation using prompts and predefined criteria (e.g., marking schemes). Compared to traditional neural models, LLMs offer the advantage of handling tasks such as question answering, reading comprehension, and summarization without requiring fine-tuning (i.e., training on task-specific datasets through supervised learning) \cite{radford2019language}. However, in automated grading, few-shot learning—supervised learning with limited training epochs—has become the standard approach. This method has been applied to state-of-the-art LLMs, including BERT \cite{chen2024multi,sung2019pre}, GPT-2 \cite{radford2019language}, GPT-3 \cite{brown2020language,cao2023leveraging,mizumoto2023exploring}, LaMDA \cite{wei2021finetuned}, and GPT-4 \cite{chiang2024large,impey2024using,xiao2024automation}. These studies demonstrate the effectiveness of LLMs in grading through benchmarking on grading datasets and evaluations conducted by educators. While there have been attempts \cite{kortemeyer2023can} to apply zero-shot learning with simple prompts on higher education courses using LLMs (i.e., without further supervised learning), it nevertheless proved inadequately effective for comprehensive assessments, such as examinations.

\textbf{Student Perceptions and Adaptive Feedback}. While LLMs have shown effectiveness in grading and supporting student learning, understanding students' perceptions on automated grading systems is equally important. Although automated grading systems have been implemented in classroom settings \cite{chen2008beyond,dikli2014automated,li2014role}, few studies have examined how students perceive these systems. Concerns may arise regarding the accuracy of automated grading, misunderstandings of system operations—resulting in suboptimal responses—and difficulties in adapting answers to align with evaluation criteria \cite{azad2020strategies,grimes2010utility,hsu2021attitudes}. Moreover, effectively identifying student mistakes remains critical, as emphasized by the LLM-based feedback tool introduced in \cite{matelsky2023large}.

\textbf{Limitations}. Although previous studies offer valuable insights into Automated Grading Systems, several limitations may hinder their further development and effectiveness:
\begin{enumerate}
    \item \textbf{Dataset Requirements}. Few-shot learning methods still require substantial datasets, especially considering the scale of LLMs. Acquiring large-scale datasets can be costly and may not be feasible for introductory courses or ad-hoc topics, where content frequently changes.  
    \item \textbf{Limited Focus on Feedback}. Automated Grading Systems often prioritize grading accuracy while neglecting additional functions such as providing meaningful feedback and comments to students.  
    \item \textbf{Benchmark-Oriented Evaluation}. These systems are typically evaluated based on their performance on grading datasets, overlooking the student experience and feedback.  
\end{enumerate}
To address these challenges, we propose a zero-shot LLM-based Automated Assignment Grading (AAG) System capable of evaluating answers that include both calculations and natural language explanations through prompt engineering. Additionally, the system provides constructive feedback to students by identifying mistakes and suggesting ways to improve. While extensive research has examined the grading effectiveness of LLMs, this study focuses on evaluating the system from students' perspectives through survey responses after grading their actual homework and delivering personalized feedback—an area that, to the best of our knowledge, remains underexplored.

\section{Zero-Shot LLM-Based Automated Assignment Grading System}

This section introduces the Zero-Shot LLM-Based Automated Assignment Grading (AAG) system. An overview of the system is illustrated in Fig. \ref{fig1}. By automatically providing feedback to both teachers and students, the AAG system could significantly reduce the manual grading workload while offering comprehensive and detailed insights to support continuous improvement. A demonstration of the AAG system is provided using a question (see Table \ref{tab:sampling_rubric}) from the STAT1011 Introduction to Statistics course assignment at The Chinese University of Hong Kong.

\subsection{Adaptive Input Processing and Marking Scheme Refinement}
The AAG system processes three key inputs: the student’s assignment submission, the assignment questions, and—optionally—a reference solution and/or marking scheme (indicated by the dashed box labeled ``non-complementary input'' in Fig. \ref{fig1}). While the reference solution and marking scheme can offer more direct guidance to the LLM, they are not essential for the grading process.

The AAG system is designed to generate or refine scoring guidelines based on the assignment questions. This feature is particularly beneficial for introductory courses like STAT1011 Introduction to Statistics, where the relevancy of assignments often depends on timely topics and current events, such as presidential elections or census surveys. This flexibility allows instructors to create more diverse and engaging questions without the need to constantly develop detailed marking schemes.

\subsection{Leveraging LLMs for Automated Grading}

The AAG system utilizes GPT-4 \cite{achiam2023gpt} for solution refinement, student submission evaluation, and summarizing student issues. Among the state-of-the-art LLMs—including Llama-3 \cite{dubey2024llama}, Qwen-2 \cite{yang2024qwen2}, Claude-3, and Gemini-1.5 \cite{team2024gemini}—in which GPT-4 was identified as the most effective model. This conclusion was based on a comparative evaluation using identical prompts, with performance assessed by a group of university lecturers.

\begin{figure}
\includegraphics[width=\textwidth]{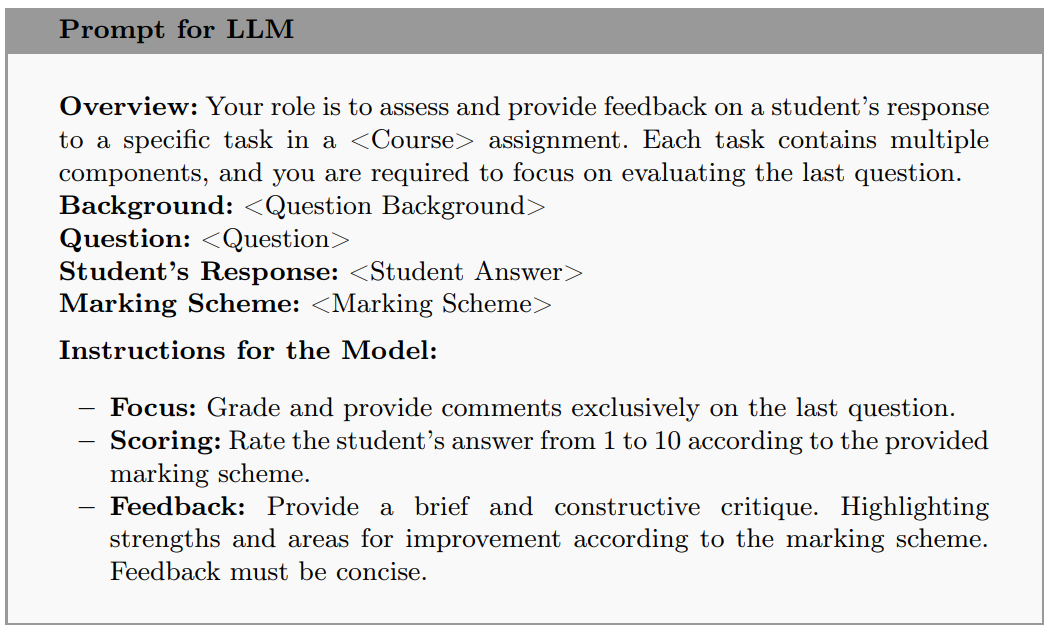}
\caption{Evaluation prompt for the AAG system. The angle brackets < > indicate placeholders for specific content. The question and answer placeholders can represent multiple questions and answers, respectively, as a single question may include several subquestions and rely on previous answers. This prompt is designed to incorporate all prior context while specifically evaluating the last included question and its corresponding answer.} \label{fig2}
\end{figure}

\begin{figure}
\includegraphics[width=\textwidth]{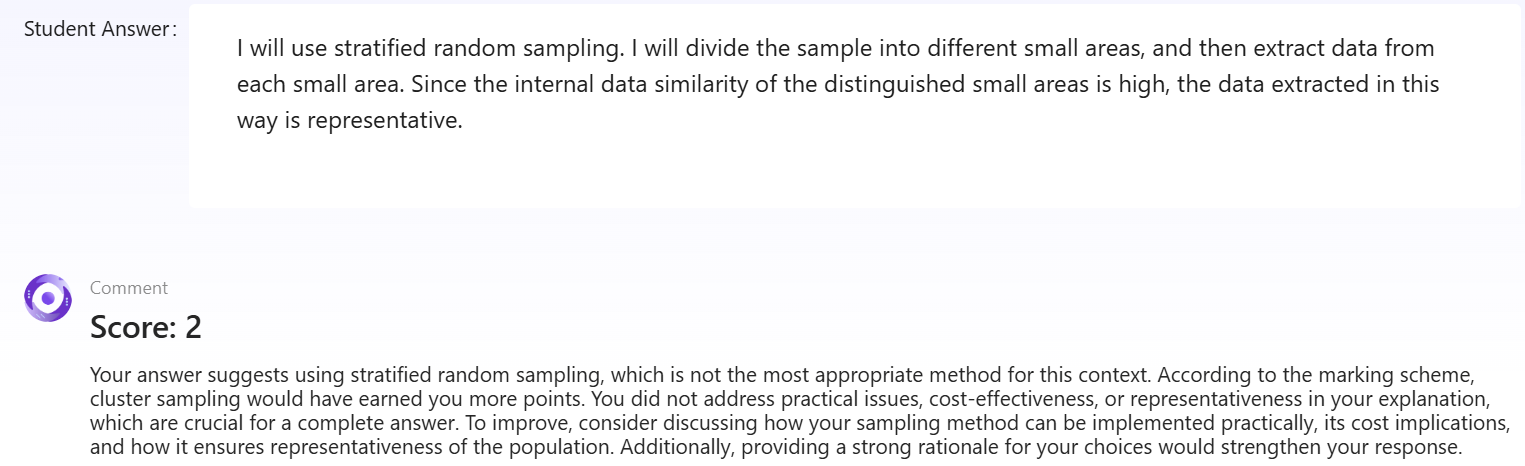}
\caption{AAG system student feedback example.} \label{fig3}
\end{figure}

The student submission, assignment questions, and marking scheme are integrated into the evaluation prompt, as illustrated in Fig. \ref{fig2}. The prompt begins by outlining the evaluation task and specifying the course context. It then provides the assignment background, question, marking scheme, student response, and correct answer (see Table \ref{tab:sampling_rubric} and Fig. \ref{fig3}). Finally, the prompt includes instructions on which specific question (or subquestion) to focus on—especially important for sequential questions—along with the scoring scale (1 to 10) and the type of feedback required for students. With a carefully engineered prompt and refined marking scheme, the LLM could provide satisfactory feedback without fine-tuning.

\subsection{Personalized Feedback for Students}

Students receive tailored feedback that includes detailed comments on their work, along with a score based on the refined marking scheme (see Fig. \ref{fig3} for an example). The feedback highlights the errors made in the student's responses, explains why those answers are incorrect, and offers suggestions for improvement. These suggestions address both the student's understanding of the material and their approach to answering questions. This type of feedback provides more actionable insights compared to traditional grading by teaching assistants, which typically only provides marks and short comments without further explanation.

\subsection{Performance Overview for Teachers}

\begin{figure}
\includegraphics[width=\textwidth]{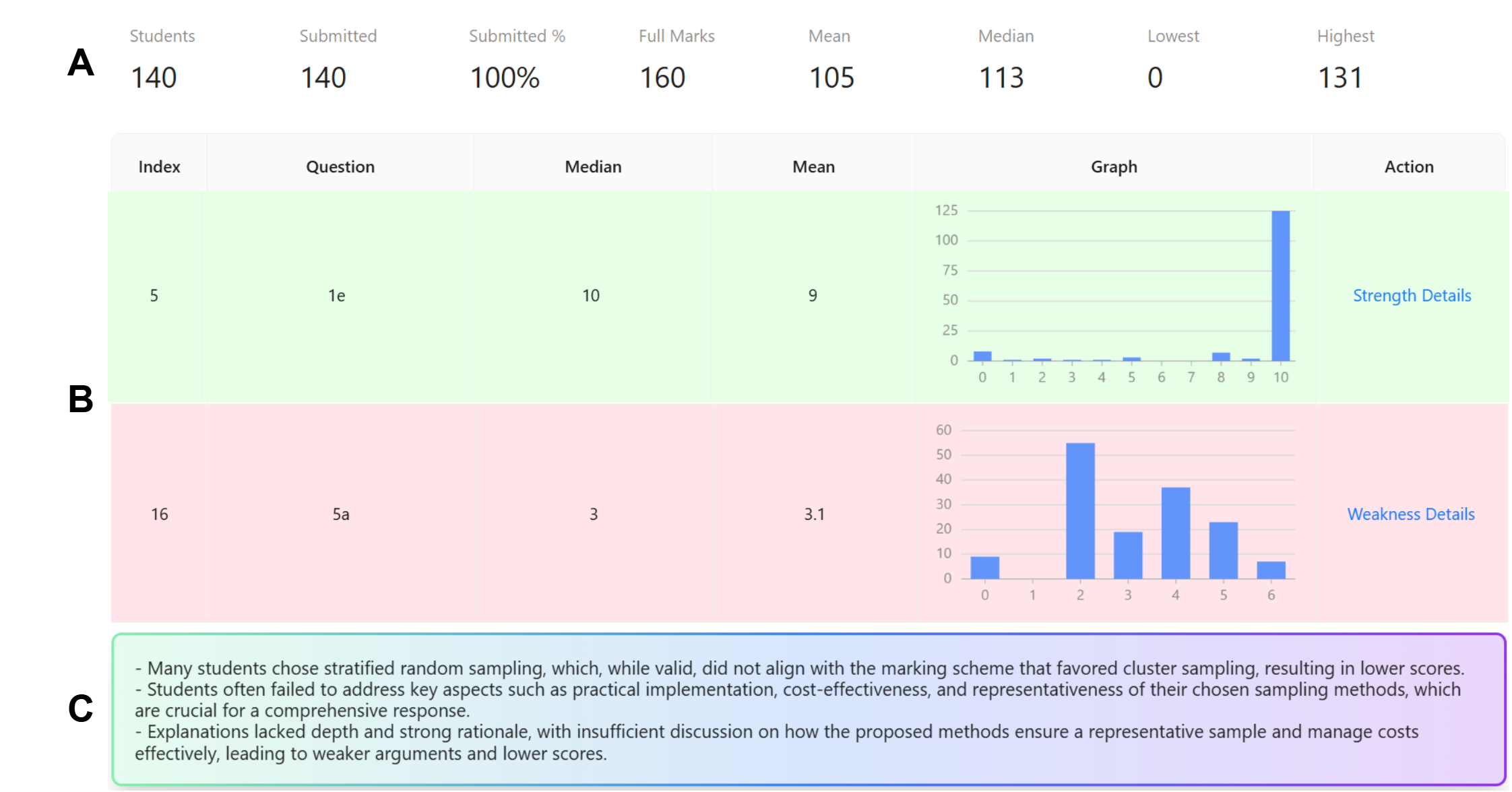}
\caption{AAG system teachers feedback example. (A) and (B) provide summaries based on the student's submission and the AAG score, while (C) presents a summary of the AAG feedback generated using LLM.} \label{fig4}
\end{figure}

For teachers, the system generates a ``Performance Summary'', providing an overall view of how the class or an individual student performed relative to the assignment objectives. An example is shown in Fig. \ref{fig4}. Fig. \ref{fig4}A presents summary statistics for the assignment, giving teachers an overview of submission and student performance. Fig. \ref{fig4}B offers a breakdown of each question in the assignment, while Fig. \ref{fig4}C illustrates an example of student issues in Question 1 (see Table \ref{tab:sampling_rubric}), it summarizes common problems based on the feedback provided to students through the LLM. This summary equips teachers with valuable insights to guide their instructional decisions and support student improvement.

\subsection{Limitations of the AAG System}
While the AAG system offers significant benefits, it faces several limitations. First, the AAG system relies on prompt engineering and marking scheme quality to ensure grading consistency and diversity. Implementing adaptive marking scheme generation could streamline the process, though human validation would still be necessary. Nevertheless, this approach would be far more efficient than traditional human grading. Second, the lack of iterative feedback restricts opportunities for student learning and growth. Introducing a multi-stage feedback mechanism in the future would enable students to revise and resubmit their work, fostering continuous improvement. Finally, limited integration with collaborative tools and resources could limit student engagement. Expanding support for widely-used collaborative platforms could enhance interaction, making the system more adaptable and fostering greater student involvement.

\section{Experiment}
This section evaluates the effectiveness of the AAG system by comparing its grading with that of human teaching assistants (TAs) on open-ended questions from the STAT1011 course. The alignment of scores between the AAG system and TAs is assessed, along with the system's impact on grading consistency and feedback quality. Additionally, student feedback from a voluntary survey is included to evaluate the system's perceived usefulness. The appendix, all data, and code for the statistical results are available at \url{https://github.com/calvinyeungck/Automated_Assignment_Grading}.

\subsection{AAG Grading Evaluation}
While there was substantial evidence demonstrating the effectiveness of LLMs in grading (see Section \ref{sec:related_work}), a limitation of LLM-based grading was the distributional differences between AI scorers and human  \cite{hashemi2024llm}. To address this issue, we compared the scores assigned by TAs and the AAG system using two open-ended questions from the STAT1011 Introduction to Statistics course. Open-ended questions were selected due to their greater flexibility in student responses compared to calculation-based questions. These two questions are summarized in Tables \ref{tab:sampling_rubric}, and \ref{tab:stratification_rubric}, respectively. For the comparison, both the TA and the AAG system independently graded 150 student assignments.

\begin{table}[h!]
\caption{Question and marking scheme for STAT1011 question 1.}
\label{tab:sampling_rubric}
\centering
\begin{tabular}{p{4.5cm}p{10cm}}
\hline
\textbf{Background} & 
You are a Statistician at the Census and Statistics Department, tasked with leading a team to collect data on the aging population in a housing estate in Shatin. The collected data will be reported to the Social Welfare Department.  
\\[8pt]  
\textbf{Question} &  
What kind of sampling method will you suggest using?  
\\ \hline
\multirow{17}{*}{\textbf{Marking Scheme}} &  
Award 3 marks if the student selects cluster sampling as the sampling method. Award 2 marks if the student proposes any other appropriate sampling method. 
\\[4pt]  
& Award 1 mark if the student addresses practical issues of the proposed sampling scheme (e.g., ease of implementation, availability of clustering/stratifying variables).  
\\[4pt]  
& Award 2 marks if the student considers the cost-effectiveness of the proposed sampling scheme.  
\\[4pt]  
& Award 1 mark for a strong, well-reasoned explanation of why the proposed method is cost-effective.  
\\[4pt]  
& Award 2 marks if the student considers the representativeness of the proposed sampling scheme.  
\\[4pt]  
& Award 1 mark for a strong, clear justification of how the proposed method ensures representativeness.  
\\ \hline
\end{tabular}
\end{table}

\begin{table}[h!]
\caption{Question and marking scheme for STAT1011 question 2.}
\label{tab:stratification_rubric}
\centering
\begin{tabular}{p{4.5cm}p{10cm}}
\hline
\textbf{Background} & 
You would like to conduct a survey to predict the results of the US presidential election.  
\\[8pt]  
\textbf{Question} &  
Find a good variable for forming strata. Explain your choice briefly.  
\\ \hline
\multirow{7}{*}{\textbf{Marking Scheme}} &  
This is an open-ended question, and the answer can be subjective.  
\\[4pt]    
& The student can choose any good variables they want. Some examples include by states and/or age group, etc.  
\\[4pt]    
& The student needs to explain that there is large between-strata variation and that a representative sample is obtained based on the chosen variable.  
\\[4pt]    
& Additionally, the student must explain that such a stratifying variable is feasible in practice.  
\\  
\hline
\end{tabular}
\end{table}

\begin{figure}
\includegraphics[width=0.5\textwidth]{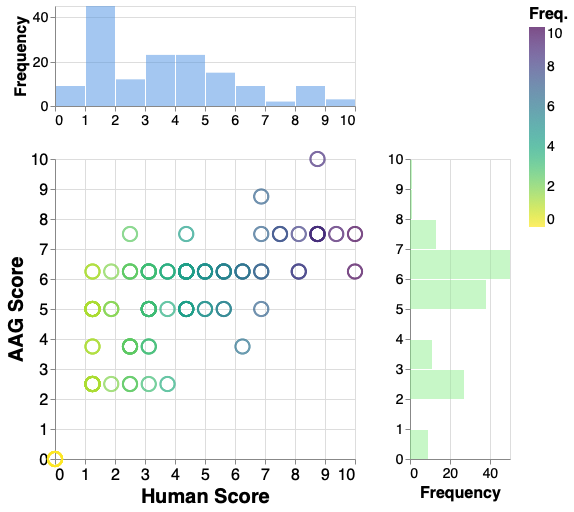}
\includegraphics[width=0.5\textwidth]{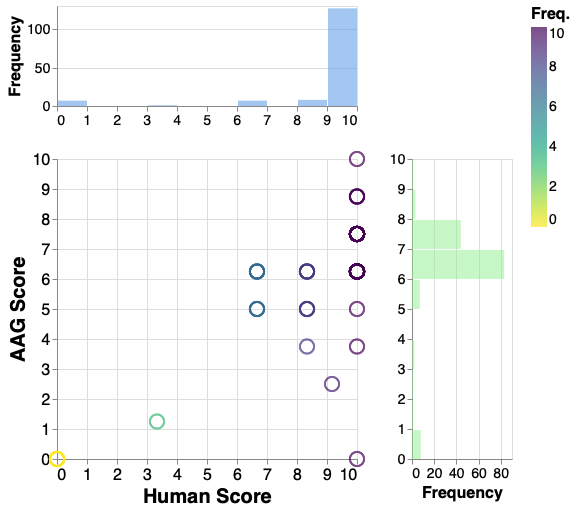}
\caption{Human and AAG system grading distribution of question 1 (left) and question 2 (right) in STAT1011.} \label{fig5}
\end{figure}

Figure \ref{fig5} presents the grading distributions of Question 1 (left) and Question 2 (right) by both human graders and the AAG system. The Pearson correlation coefficients between human and AAG scores for Questions 1 and 2 were 0.75 and 0.82, respectively. These strong correlations indicate that the AAG system produced grading results closely aligned with human evaluations, indirectly supporting the grading accuracy of the AAG system (direct validation against ground truth was not possible due to the absence of reference grading data, which would serve as the ground truth).

Despite the high correlations, noticeable differences in score distributions were observed between the two questions. Although both were open-ended, the marking schemes differed significantly. The marking scheme for Question 1 (see Table \ref{tab:sampling_rubric}) provided detailed guidelines for awarding marks, whereas the marking scheme for Question 2 (see Table \ref{tab:stratification_rubric}) offered only general grading criteria. As a result, both human and AAG scores for Question 1 were more diverse, while scores for Question 2 were more concentrated. This highlights the importance of the refined marking scheme function, which ensures more consistent and precise grading.

A further qualitative analysis of scoring discrepancies across all assignment questions revealed that some differences stemmed from human factors, such as grading errors (e.g., overlooking a missing zero, incorrect calculation processes, or missing explanations). These inconsistencies often arose when TAs failed to strictly follow the marking scheme, encountered vague grading guidelines, or made errors due to the large volume of assignments. These analysis highlights the importance of implementing the AAG system to enhance grading consistency and provide higher-quality feedback to students.

\subsection{Student Perspective on AAG System}

\textbf{Survey Design:} To evaluate the effectiveness of the AAG system, a voluntary survey was administered to students who received AAG system feedback. Participants included students from the undergraduate course STAT 1011: Introduction to Statistics and the graduate courses STAT 5206: Data Management and RMSC 5002: Principles of Risk Management at The Chinese University of Hong Kong. Participants were informed that their responses would not impact their assignment scores. The survey included 10 questions aimed at assessing the quality of the system's feedback and its influence on learning. Questions 1 to 9 employed a 5-point Likert scale (1 $=$ Strongly Disagree, 5 $=$ Strongly Agree), while Question 10 asked students to choose between the traditional TA grading method and the AAG system feedback.

\textbf{Survey Questions:} The survey questions targeted multiple dimensions of the feedback experience (The exact survey question is available in Section \ref{sec:survey_quesiton}):

\begin{itemize}
\item \textbf{Q1-Q2:} Focused on the clarity and helpfulness of the feedback in identifying specific mistakes.
\item \textbf{Q3-Q4:} Evaluated how well the feedback clarified why answers were incorrect and the degree of customization to individual responses.
\item \textbf{Q5-Q6:} Measured how effectively the feedback provided actionable steps for improvement and addressed personal learning needs.
\item \textbf{Q7-Q8:} Assessed the impact of feedback on future problem-solving and its insight into knowledge gaps.
\item \textbf{Q9:} Investigated whether the feedback encouraged motivation to learn and correct mistakes.
\item \textbf{Q10:} Asked students to choose between conventional TA grading and the AAG system feedback.
\end{itemize}

\begin{figure}
\includegraphics[width=0.55\textwidth]{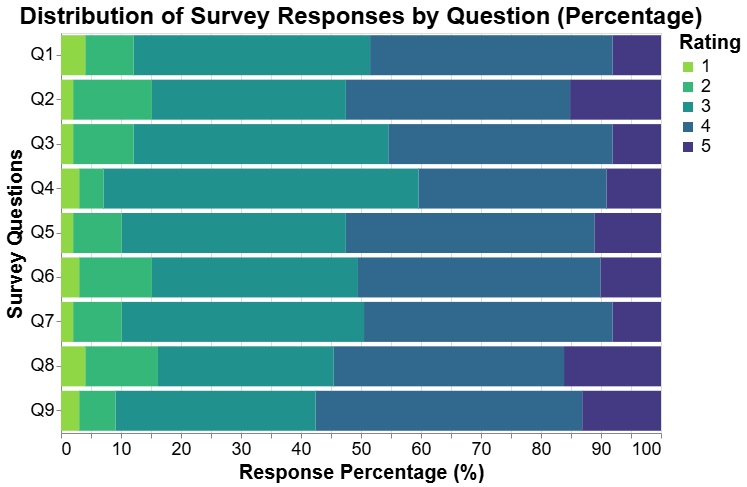}
\includegraphics[width=0.45\textwidth]{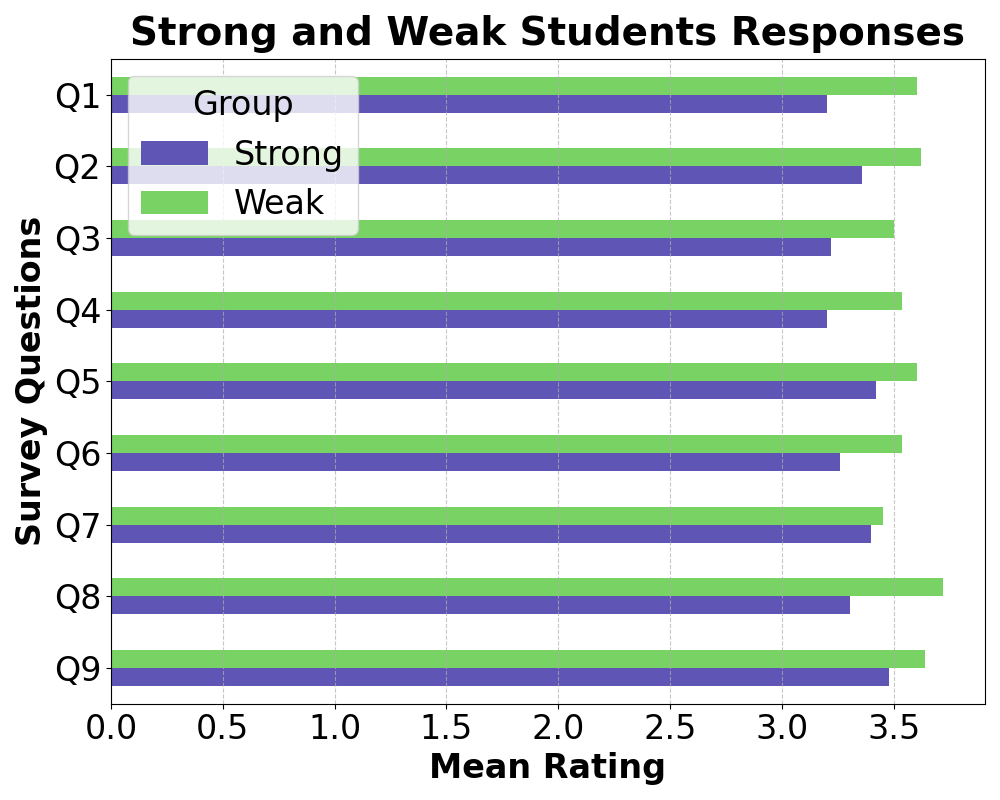}
\caption{Distribution of survey responses. For questions 1 through 9, responses follow a Likert scale where 5 indicates the highest rating and 1 indicates the lowest. The left plot displays the distribution of responses from all students, while the right plot shows the mean ratings for strong and weak students (defined as the top 50\% and bottom 50\% performers, respectively).} \label{fig6}
\end{figure}

\textbf{Results and Interpretation:} In total 104 responses were collected and the survey results are summarized in Fig. \ref{fig6} (left). A Wilcoxon Signed-Rank Test and a Binomial Test were conducted on the responses to Questions 1-9 and Question 10, respectively (detailed results can be found in Section \ref{sec:h_test}). With strong statistical significance (p < 0.0001), the tests revealed consistently positive responses across all survey items. Specifically, for Questions 1-9, students indicated that the AAG system significantly improved their understanding of mistakes, provided clearer explanations, and delivered highly customized feedback tailored to their learning needs. Moreover, the feedback was effective in offering actionable steps for improvement, thereby enhancing students' motivation to learn.

Specifically, the survey results indicated that students felt more prepared to tackle similar problems in the future due to the AAG system’s feedback, which offered deeper insights into their learning gaps. The system's ability to provide personalized feedback contributed to a more engaging and effective learning experience. For Question 10, the majority of students (93 out of 104) preferred the AAG system over traditional TA grading, highlighting its perceived superiority in offering constructive, individualized feedback that better supports student learning.

Furthermore, the responses from weak and strong student groups, defined as the top 50\% and bottom 50\% performers on the corresponding assignment, were compared to identify any significant differences in perception or understanding between the two groups. As shown in Figure \ref{fig6} (right), the mean ratings indicate that weak students generally gave higher ratings to the AAG system across all aspects.

A Mann-Whitney U test\footnote{Note that the power of the parametric t-test and the non-parametric Mann-Whitney-Wilcoxon test are similar for five-point Likert scale data \cite{de2010five}.} was also performed to assess the significance of these differences (detailed results in Section \ref{sec:h_test_group}). With strong confidence (p < 0.05), the analysis revealed that weaker students were more satisfied with the AAG system in four key areas.

For Q1 (Identifying Mistakes), weaker students found the detailed feedback especially helpful in pinpointing mistakes. For Q4 (Actionable Feedback for Improvement), they also felt the system guided them better toward improvement strategies. For Q6 (Future Problem-Solving), they felt more prepared for future problems, enhancing their confidence and skills. Lastly, for Q8 (Encouragement for Learning), the system effectively motivated and encouraged continuous learning, especially for struggling students.

Overall, these findings highlight the AAG system's substantial impact on enhancing feedback quality, student engagement, and learning outcomes. The system was particularly effective in supporting weaker students by improving their ability to identify mistakes, providing practical guidance for improvement, boosting their confidence in problem-solving, and fostering greater motivation to learn. These results underscore the value of personalized, actionable feedback in addressing the specific needs of students, especially those who may struggle, ultimately contributing to a more inclusive and effective learning environment.

\section{Conclusion}
This paper presents an Assignment Assessment and Grading (AAG) system that uses a zero-shot LLM framework to evaluate assignments effectively without requiring additional training and datasets. The system delivers personalized feedback, helping students identify strengths and areas for improvement, thus enhancing learning outcomes.  In comparison with human grading using actual student assignments, the AAG system could provide more consistent and higher-quality feedback. Additionally, survey results from students affirm that AAG system-tailored feedback could significantly enhance motivation, understanding, and preparedness, outperforming traditional grading methods.

Future work will focus on expanding the system to integrate with Learning Management Systems for real-time and multi-stage feedback and adapting it to other academic disciplines. Additionally, exploring multimodal feedback, conducting longitudinal studies on the system’s impact, and developing personalized learning paths are key areas of development. Ethical considerations, including addressing bias, will be prioritized to ensure equitable and fair feedback for all students.

\begin{credits}
\subsubsection{\ackname} This study was funded by the Hong Kong University Grants Committee (UGC) Fund for Innovative Technology-in-Education (FITE), 2023-2026. The authors extend their sincere gratitude to Dr. Issac Leung, Chan Shun Wai, Li You Tong, Qi Zi Han, and the other teaching assistants for their valuable assistance throughout this study.

\subsubsection{\discintname}
The authors have no competing interests.
\end{credits}

%
%
%
\bibliographystyle{splncs04}
\bibliography{ref}
%





\newpage
\begin{appendix}

\section{Hypothesis Testing for Survey Responses}
\label{sec:h_test}

This section outlines the hypothesis testing performed on the survey responses, using statistical tests to evaluate the significance of the responses for each question.

\noindent \textbf{1. Wilcoxon Signed-Rank Test} (for Q1 to Q9)
\begin{itemize}
    \item \textbf{Null Hypothesis ($H_0$):} The median response for each question (Q1 to Q9) is less then or equal to 3.
    \item \textbf{Alternative Hypothesis ($H_1$):} The median response for each question (Q1 to Q9) is greater than 3.
\end{itemize}
\textbf{2.Binomial Test} (for Q10)
\begin{itemize}
    \item \textbf{Null Hypothesis ($H_0$):} The proportion of binary responses is less than or equal to 0.5.
    \item \textbf{Alternative Hypothesis ($H_1$):} The proportion of binary responses greater than 0.5.
\end{itemize}
The hypotheses are determined based on the following conditions:
\begin{itemize}
    \item If the p-value is less than 0.05, the null hypothesis will be rejected, indicating a significant difference.
    \item If the p-value is greater than 0.05, the null hypothesis cannot be rejected, suggesting no significant difference.
\end{itemize}

The results of the hypothesis tests for questions Q1 to Q9 (Wilcoxon Signed-Rank Test) and Q10 (Binomial Test) are presented in Table \ref{tab:wilcoxon_binom_test}.

\begin{table}[ht]
\centering
\caption{Wilcoxon Signed-Rank test result for Q1-Q9 and Binomial test result for Q10.}
\begin{tabular}{|c|c|c|c|c|}
\hline
\textbf{Question} & \textbf{Test Statistic} & \textbf{p-value} & \textbf{Sample Size} & \textbf{Interpretation ($\alpha$=0.05)} \\
\hline
Q1 & 1523  & < 0.0001  & 103 & Significant  (greater than 3) \\
Q2 & 1888  & < 0.0001 & 103 & Significant  (greater than 3) \\
Q3 & 1335  & < 0.0001 & 104 & Significant  (greater than 3) \\
Q4 & 1000  & < 0.0001 & 104 & Significant  (greater than 3) \\
Q5 & 1794  & < 0.0001 & 103 & Significant  (greater than 3) \\
Q6 & 1827  & < 0.0001 & 104 & Significant  (greater than 3) \\
Q7 & 1490  & < 0.0001 & 103 & Significant  (greater than 3) \\
Q8 & 2131  & < 0.0001 & 103 & Significant  (greater than 3) \\
Q9 & 1883  & < 0.0001 & 103 & Significant  (greater than 3) \\
Q10 & 0.8942  & < 0.0001 & 104 & Significant  (greater than 0.5) \\
\hline
\end{tabular}
\label{tab:wilcoxon_binom_test}
\end{table}

For all questions Q1 to Q9, the p-values are less than 0.05, indicating that the null hypothesis is rejected, and the median response for each question is significantly greater than 3. This suggests that, on average, the respondents rated each of these questions more favorably than the neutral midpoint (3). For Q10, the p-value is also less than 0.05, leading to the rejection of the null hypothesis. This means that the proportion of positive binary responses is significantly greater than 0.5, indicating a preference for the positive response.

\section{Hypothesis Testing for Between-Group Comparison}
\label{sec:h_test_group}

This section describes the use of the \textbf{Mann-Whitney U Test} (for Q1 to Q10) for comparing the survey responses between two groups of students: weak students (Group 1) and strong students (Group 2). For the comparison between weak and strong students, the hypotheses are as follows:
\begin{itemize}
    \item \textbf{Null Hypothesis ($H_0$):} There is no significant difference between the weak students (Group 1) and the strong students (Group 2) with respect to the survey responses, i.e., the distribution of survey responses for weak students is less than or equal to that of strong students.
    \item \textbf{Alternative Hypothesis ($H_1$):} The distribution of survey responses for weak students is greater than that of strong students.
\end{itemize}
The hypotheses are determined based on the following conditions:
\begin{itemize}
    \item If the p-value is less than 0.05, the null hypothesis will be rejected, indicating a significant difference.
    \item If the p-value is greater than 0.05, the null hypothesis cannot be rejected, suggesting no significant difference.
\end{itemize}

The results of the Mann-Whitney U test comparing weak and strong students (Groups 1 and 2) for each survey question are summarized in Table \ref{tab:group_h_test_result}. The interpretation of each result follows the decision rule outlined above.

\begin{table}[!ht]
\centering
\caption{Comparison between weak and strong students (group 1 and group 2) with interpretations. Where n denotes the sample size of the respective group.}
\begin{tabular}{|c|c|c|c|c|c|}
\hline
\textbf{Survey Question} & \textbf{Test Statistic} & \textbf{p-value} & \textbf{n$_{group 1}$} & \textbf{n$_{group 2}$} & \textbf{Interpretation ($\alpha$=0.05)} \\ \hline
Q1 & 1687 & 0.0053 & 53 & 50 & \text{Significant } \\ 
Q2 & 1537 & 0.0717 & 53 & 50 & \text{Not Significant } \\ 
Q3 & 1561 & 0.0711 & 54 & 50 & \text{Not Significant } \\ 
Q4 & 1603 & 0.0350 & 54 & 50 & \text{Significant } \\ 
Q5 & 1482 & 0.1355 & 53 & 50 & \text{Not Significant } \\ 
Q6 & 1594 & 0.0471 & 54 & 50 & \text{Significant } \\ 
Q7 & 1354 & 0.4212 & 53 & 50 & \text{Not Significant } \\ 
Q8 & 1655 & 0.0109 & 54 & 49 & \text{Significant } \\ 
Q9 & 1515 & 0.0905 & 53 & 50 & \text{Not Significant } \\ 
Q10 & 1335 & 0.5751 & 54 & 50 & \text{Not Significant } \\ \hline
\end{tabular}
\label{tab:group_h_test_result}
\end{table}

From the Mann-Whitney U test results, significant differences between weak and strong students were observed for Q1, Q4, Q6, and Q8, where the p-values were less than 0.05. This indicates that the survey responses of weak students were significantly higher than those of strong students for these questions. For the remaining questions (Q2, Q3, Q5, Q7, Q9, and Q10), the p-values exceeded 0.05, suggesting no significant differences in the responses between the two groups.

\section{Survey Questions}
\label{sec:survey_quesiton}
This section includes the set of survey questions designed to gather valuable insights into various aspects of the AAG system. The goal is to understand students' perspectives, experiences, and feedback, which will inform future development and improvements.



\section*{Supplementary material}

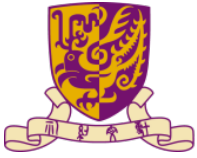

**Evaluating the Impact of AI-Generated Feedback on Student Learning**

This survey aims to evaluate an AI tool that provides personalized feedback on student assignments. By comparing the standard solution with AI-generated feedbacks tailored to your answers, we hope to understand how this comment impacts learning, especially for those needing extra guidance.

Your responses will help us improve the tool to better support students' understanding.

* Your SID：

Q1 Identifying Mistakes

To what extent did the AI feedback help you identify the specific mistakes in your answers, compared to just seeing the solution alone?

○ (1) No additional help in identifying mistakes

○ (2) Very little additional help

○ (3) Some additional help

○ (4) Noticeable improvement in identifying mistakes

○ (5) Much clearer identification of mistakes

**Q2 Clarity of Explanation for Errors**

Did the AI comment clarify why your answers were incorrect better than the standard solution alone could?

○ (1) No added clarity

○ (2) Slightly clearer but limited

○ (3) Moderately clearer

○ (4) Significantly clearer explanation

○ (5) Very clear and detailed explanation

**Q3 Level of Customization**

How well was the AI feedback tailored to your specific answer compared to a general solution?

○ (1) Not customized at all

○ (2) Minimally customized

○ (3) Somewhat customized

○ (4) Well-customized

○ (5) Very highly customized to my answer

**Q4 Benefit of Customization**

How much did the customized AI comment, tailored to your specific answer, help you address your personal learning needs compared to a general solution?

○ (1) No additional help for my learning needs

○ (2) Very minimal additional help

○ (3) Somewhat helpful for my needs

○ (4) Noticeably helpful for my needs

○ (5) Extremely helpful and targeted to my learning needs

**Q5 Actionable Feedback for Improvement**

How effective was the AI feedback in providing specific steps or areas to improve, compared to the solution alone?

○ (1) Not effective in providing improvement steps

○ (2) Very limited effectiveness

○ (3) Somewhat effective

○ (4) Effective in providing actionable steps

○ (5) Extremely effective with clear, actionable feedback

Next page ›

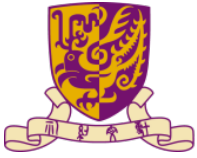

**Q6 Future Problem-Solving**

How prepared do you feel to handle similar problems in the future with AI feedback, as opposed to only seeing the solution?

○ (1) Not more prepared than with solution alone

○ (2) Slightly more prepared

○ (3) Moderately more prepared

○ (4) Significantly more prepared

○ (5) Very confident in handling similar problems

**Q7 Depth of Insight into Learning Gaps**

To what extent did the AI feedback help you identify specific knowledge gaps, in contrast to the general solution alone?

○ (1) No additional insight into my learning gaps

○ (2) Minimal additional insight

○ (3) Some additional insight

○ (4) Clearer insight into learning gaps

○ (5) Provided significant, valuable insight

**Q8 Encouragement for Learning**

How much more motivated did you feel to address your mistakes after receiving the AI comment, as compared to just the solution?

○ (1) Not more motivated than with solution alone

○ (2) Slightly more motivated

○ (3) Somewhat more motivated

○ (4) Noticeably more motivated to improve

○ (5) Very motivated to address and learn from mistakes

**Q9 Comparative Learning Benefit**

To what extent did the AI feedback contribute to an improved learning experience compared to receiving only the solution?

○ (1) No additional benefit compared to solution alone

○ (2) Slight additional benefit

○ (3) Moderate additional benefit

○ (4) Significant benefit for learning

○ (5) Greatly enhanced learning experience

**Q10 Preference Between Solution Types**

Overall, which did you find more helpful for understanding: the solution alone or the solution with the AI feedback?

○ (1) Solution alone,

○ (2) Solution + AI Comment

**Q11 Suggestions for AI Feedback Improvement (Open-Ended)**

What additional information or aspects in the AI feedback would make it even more helpful compared to the solution alone?

Next page ❯








\end{appendix}

\end{document}